\documentclass[onecolumn,preprintnumbers,amssymb,amsmath,superscriptaddress,letterpaper,nofootinbib]{revtex4}[12pt]
\usepackage{times,graphics}
\usepackage[colorlinks]{hyperref}
\usepackage{makecell}
\usepackage{diagbox}

\usepackage{graphicx}
\usepackage{dcolumn}
\usepackage{bm}
\usepackage{natbib}
\usepackage{graphicx}
\usepackage{epsfig}
\usepackage{epstopdf}
\usepackage{multirow}
\usepackage{amsmath}
\usepackage{lipsum}

\newcommand{\physrep}{{Physics~Reports}}

\newcommand{\apjl}{{Astrophys.~J.~Lett.}}
\newcommand{\apjs}{{Astrophys.~J.~Supp.}}
\newcommand{\aj}{{Astron.~J.}}
\newcommand{\mnras}{{Mon.~Not.~R.~Astron.~Soc.}}

\newcommand{\be}{\begin{equation}}
\newcommand{\ee}{\end{equation}}
\newcommand{\bea}{\begin{eqnarray}}
\newcommand{\eea}{\end{eqnarray}}

\makeatletter
\newcommand*{\shifttext}[2]{%
	\settowidth{\@tempdima}{#2}%
	\makebox[\@tempdima]{\hspace*{#1}#2}%
}
\makeatother

\graphicspath{{./fig/}}

\begin{document}

	\title{Does $\Lambda$CDM really be in tension with the Hubble diagram data?}

	\author{Ahmad Mehrabi}
	
	\affiliation{Department of Physics, Bu-Ali Sina University, Hamedan
		65178, 016016, Iran}
	\email{mehrabi@ipm.ir}
	
	\author{Spyros Basilakos}
	\affiliation{Academy of Athens, Research Center for Astronomy \& Applied
		Mathematics, Soranou Efessiou 4, 11-527, Athens, Greece}
	\email{svasil@academyofathens.gr}
	\date{\today}

	\begin{abstract}
		In this article, we elaborate further on the  
		$\Lambda$CDM ''tension'', suggested recently by the authors  
		\cite{Lusso:2019akb,Risaliti:2018reu}. 
		We combine Supernovae type Ia (SNIa) with quasars (QSO) and Gamma Ray Bursts (GRB) 
		data in order to reconstruct
		in a model independent way the Hubble 
		relation to as high redshifts as possible.
		Specifically, in the case of either SNIa or SNIa/QSO data 
		we find that current values of  the cosmokinetic parameters extracted from the  
		Gaussian process are consistent with those of $\Lambda$CDM. 
		Including GRBs in the analysis we find a tension, which however
		is not as significant as that mentioned in 
		\cite{Lusso:2019akb,Risaliti:2018reu}.
		Finally, we argue that 
		that the choice of the kernel function used in extracting the 
		luminosity distance might affect the amount of tension. 
		
	\end{abstract}
	\maketitle

\section{Introduction}
Since the discovery of the accelerated expansion of the Universe from 
the Supernovae type Ia (SNIa) 
data\cite{Riess1998,Perlmutter1999}, the combined
analysis of various cosmological probes, including those of 
Cosmic Microwave Background (CMB) \cite{Komatsu2011,Ade:2015yua,Aghanim:2018eyx}, Baryon Acoustic Oscillation (BAO) \cite{Eisenstein:2005su,Percival2010,Blake:2011rj,Reid:2012sw,Abbott:2017wcz,Alam:2016hwk,Gil-Marin:2018cgo} and cosmic chronometers \cite{Farooq:2016zwm} confirms the aforementioned 
dynamical result, namely that currently the 
Universe accelerates. However, the physics of cosmic acceleration is still a 
mystery, hence the aim in these kind of studies is to 
provide an explanation regarding the underlying mechanism which triggers 
such a phenomenon.
   
In the framework of homogeneous and isotropic Universe, the accelerated 
expansion can be described by considering either an exotic matter 
with negative pressure \cite{Weinberg:1989,Peebles:2002gy,Copeland:2006wr,Chiba:2009nh,Amendola:2010,Mehrabi:2018dru,Mehrabi:2018oke} or a modification 
of gravity \cite[$f(R)$ theories and the like,][]{Schmidt:1990gb,Magnano:1993bd,Dobado:1994qp,Capozziello:2003tk,Carroll:2003wy}. 
Among the large family of dark energy and modified gravity models, 
the simplest case is the spatially flat $\Lambda$CDM model for which
cold dark matter (CDM) and baryonic matter coexist with the
cosmological constant. From the theoretical viewpoint,
the $\Lambda$CDM model suffers from the well known problems, namely
the coincidence and the expected value of the vacuum energy density  
\cite{1989RvMP...61....1W,2003PhR...380..235P,perivolaropoulos2008puzzles,padilla2015lectures}. 

On the other hand, despite the fact that the $\Lambda$CDM model is found to be 
in a very good agreement with the majority of cosmological 
data~\cite{Aghanim:2018eyx}, nonetheless the model seems 
to be currently in tension with some recent measurements~\cite{Verde_2019,Sola:2017znb,2019PhRvD.100b3539R,2019ApJ...886L...6S}, 
related with the Hubble constant $H_0$ and 
the present value of the mass variance at 8$h^{-1}$Mpc, namely $\sigma_8$. 
Moreover, Lusso {\it et al.} \cite{Lusso:2019akb} using 
a combined Hubble diagram of SNIa, Quasars, and gamma-ray bursts (GRBs) 
found a $\sim 4\sigma$ tension between the 
best fit cosmographic parameters with respect to those of $\Lambda$CDM
(see also \cite{Risaliti:2018reu,PhysRevD.101.043502}).  In the light
of the latter results, a heated debate is taking place in the literature
and the aim of the present article is to contribute to this debate.

Here, we focus on a model-independent 
parametrization of the Hubble diagram using the Gaussian process, 
and investigate its performance against the latest Hubble diagram 
data. Notice that in this case we need to introduce 
a kernel function with some hyperparameters which can be optimized in order 
to fit the data. For more details concerning 
model-independent methods we refer to  
\cite{Liao:2019qoc,Zhang:2018gjb,Gomez-Valent:2018hwc,Melia:2018tzi}. 
The structure of the paper is as follows. 
In section  \ref{sec:gau}, we introduce the concept of 
the Gaussian process and we present the corresponding 
kernel functions that we shall use in the current work.
In section \ref{sec:data}, we discuss the observational data 
and the procedure of our analysis, while in section \ref{sec:res}
we provide our results. Finally, 
in \ref{sec:con},
we summarize our results and we draw our conclusions.

\section{Model independent method- Gaussian process}\label{sec:gau}
We consider that the universe is a
self-gravitating fluid, endowed with a spatially flat homogeneous and 
isotropic geometry. In this context, there are two main approaches in order to investigate 
cosmological data e.g. the luminosity distance. In the first case we impose a 
cosmological model, hence we estimate the form of the luminosity distance.
Then we fit the model to data in order to place constraints on the 
corresponding parameter space. This is a model-depended method is a sense 
that different models provide different forms of luminosity distance.
Another avenue is to utilize a model independent method in reconstructing
the Hubble diagram through the observational data
\cite{Liao:2019qoc,Zhang:2018gjb,Gomez-Valent:2018hwc,Melia:2018tzi}. 
In this approach we do not need to know apriori the underlying 
cosmological model. 
One of the most popular model independent method is the Gaussian process (GP),
hence in the present article we test the performance of 
GP against the available Hubble diagram data. 

Briefly, the main steps of the method are the following.
Having a data set $D$
\begin{equation}\label{eq:data-set}
	D=\{(x_i,y_i)|i=1,..,n\},
\end{equation}
our aim is to reconstruct in a model independent way 
a function $f(x)$ which describes the data. 
In this case at any point $x$, the value $f(x)$ is a 
Gaussian random variable with mean $\mu(x)$ and variance $Var(x)$. 
Moreover, the function values at any two different points are not 
independent from each other, hence the 
covariance function $cov(f(x),f(\tilde{x}))=k(x,\tilde{x})$ 
describes the corresponding correlations. 
Therefore, having an observational data set $(x_i,y_i)$ and considering 
a kernel function $k(x,\tilde{x})$, it is straightforward to 
compute the value of function and its covariance 
(for more detail see \cite{Seikel:2012uu}). 
Concerning the functional form 
of the kernel, 
there is a wide range of possibilities. In the current work
we restrict our analysis to the following parametrizations:

\begin{equation}\label{eq:cov-squ}
k(x,\tilde{x}) = \sigma_f^2\exp(-\frac{(x-\tilde{x})^2}{2l^2}),
\end{equation}
\begin{equation}\label{eq:cov-m72}
k(x,\tilde{x}) =  \sigma_f^2\exp(-\sqrt{7}\frac{|x-\tilde{x}|}{l})(1+\sqrt{7}\frac{|x-\tilde{x}|}{l}+14\frac{(x-\tilde{x})^2}{5l^2}+7\sqrt{7}\frac{|x-\tilde{x}|^3}{15l^3}),
\end{equation}  
and 
  \begin{equation}\label{eq:cov-m92}
  k(x,\tilde{x}) =  \sigma_f^2\exp(-3\frac{|x-\tilde{x}|}{l})(1+3\frac{|x-\tilde{x}|}{l}+27\frac{(x-\tilde{x})^2}{7l^2}+18\frac{|x-\tilde{x}|^3}{7l^3}+
  27\frac{(x-\tilde{x})^4}{35l^4}).
  \end{equation} 
Notice that (\ref{eq:cov-m72}) and  (\ref{eq:cov-m92}) are the 
so called Matern ($\nu=7/2$ and $\nu=9/2$) 
formulas respectively. 
 It is worth noting that the family of Matern kernels is 
a generalization of kernel (\ref{eq:cov-squ}) 
and it is widely used in multivariate statistical analysis.
In this case the absolute exponential kernel is parameterized 
by an additional parameter $\nu$. If $\nu$ goes to infinity then the 
kernel reduces to Eq.(\ref{eq:cov-squ}), while in the 
case of $\nu=1/2$ the kernel 
becomes equivalent to the absolute exponential kernel.
Also $\sigma_f$ and $l$ are two hyperparameters 
which can be constrained from the observational data.     
Since the kernel function plays a role in reconstructing $f(x)$
(in our case comoving distance), we have decided 
to use the aforementioned kernels in order to test whether 
the choice of the kernel can affect the amount of the so called
$\Lambda$CDM cosmokinetic tension.


Here we use the GAPP code \cite{Seikel:2012uu} in order to reconstruct 
$f(x)$ and its derivatives. Specifically, $f(x)$ and its derivatives are given by
     \begin{eqnarray}\label{eq:f-der}
  &f(x)& \sim GP(\mu(x),k(x,\tilde{x}))\\
  &f'(x)& \sim GP(\mu'(x),\frac{\partial^2k(x,\tilde{x})}{\partial x \partial \tilde{x}})\\
  &f''(x)& \sim GP(\mu''(x),\frac{\partial^4k(x,\tilde{x})}{\partial^2 x \partial^2 \tilde{x}}),
  \end{eqnarray}  
  where $GP$ stands for Gaussian process.

\section{Observational data and method}\label{sec:data}
The luminosity distance is the ideal tool 
to investigate the Hubble diagram. Our aim is to extend the Hubble 
relation to as high redshifts as possible, hence 
in addition to SNIa, we also consider QSOs and GRBs.
In particular, bellow we briefly present the type of standard candles, 
used in the statistical analysis.

\begin{itemize}
	\item \textit{Supernovae (SNIa)}: we utilize the ``Pantheon'' 
compilation of SNIa data \citep{Scolnic:2017caz}. 
This sample contains 
1048 spectroscopically confirmed SNIa in the redshift range $0.01<z<2.26$.

	\item \textit{Quasars (QSOs)}: Furthermore, we use 
 the sample of 1598 QSOs as collected by 
\cite{Risaliti:2015zla,Risaliti:2018reu}.
The redshift interval of the current data is $0.04<z<5.1$. Notice that, in our analysis we use bin-averaged version of QSOs data.

	\item In addition to the above data, we use a compilation of 162 GRBs
\cite{Demianski:2016dsa,Demianski:2016zxi,Amati:2013sca}
in the range of $0.03<z<9.3$.
Unlike SNIa, QSOs and GRBs 
are observed up to very high redshifts ($z>3$)
at which the distance
modulus is more sensitive to the cosmological parameters
\cite{Plionis:2011jj}.
\end{itemize}

The evolution of the distance modulus is given 
by $\mu(z)=5{\rm log}D_{L}(z)+25$, hence 
      \begin{equation}\label{eq:mu-lom1}
      D_L(z) = 10^{(\mu(z) - 25)/5},
      \end{equation} 
where $D_{L}(z)$ is the luminosity distance from which the 
normalized comoving distance
\footnote{For the rest of the paper 
$D(x)$ plays the role of $f(x)$.}
is written as
   \begin{equation}\label{eq:mu-lom2}
 D(z) = \frac{H_0}{c} \frac{D_L(z)}{1+z}.
   \end{equation}
Notice that $H_0$ is the Hubble constant and $c$ is the speed of light.

Based on the above, we compute the normalized comoving distance data points
and then we use them in order to reconstruct the form of 
$D(z)$ as well as its derivatives. 
As a matter of fact knowing $D(z)$ and its derivatives, it is 
straightforward to compute the Hubble function $H(z)$ as well as 
its first and second derivatives, namely 
    \begin{eqnarray}\label{eq:conv-H}
 &H(z)&  = \frac{H_0}{D'(z)}, \\
 &H'(z)& = -H_0 \frac{D''(z)}{D'(z)^2}, \\
 &H''(z)& = H_0[\frac{2D''(z)^2}{D'(z)^3}-\frac{D'''(z)}{D'(z)^2}].
   \end{eqnarray}
Moreover using the error propagation we obtain 
   \begin{eqnarray}\label{eq:H-error}
 &\delta H(z)& = H_0 \frac{\delta D'(z)}{D'(z)^2},\\
 &\delta H'(z)&  = H_0[\frac{\delta D''(z)}{D'(z)^2} - \frac{2D''(z)\delta D'(z)}{D'(z)^3}],\\
 &\delta H''(z)& = H_0 [\frac{\delta D'''(z)}{D'(z)^2} - \frac{2D'''(z)\delta D'(z)}{D'(z)^3} - \frac{4D''(z)\delta D''(z)}{D'(z)^3}+\frac{6D''(z)^2\delta D'(z)}{D'(z)^4}].
 \end{eqnarray}
Notice that in above formula, we use the same $H_0$ which has been used to obtain the normalized distance in Eq.(\ref{eq:mu-lom2}) and for those quantities with more than one term in uncertainty, we use square root of all terms. For example, for $\delta X = \delta a + \delta b + \delta c +... $, the total uncertainty is $\delta X = \sqrt{(\delta a)^2 + (\delta b)^2 +(\delta c)^2 +...}$.

Following the same notations we compute 
the deceleration and jerk parameters as well as the corresponding
uncertainties. As a function of $D(z)$, these parameters are:
    \begin{eqnarray}\label{eq:q-j-d}
 &q(z)& =-(1+z)\frac{D''(z)}{D'(z)} - 1,\\
 &j(z)& = (1+z)^2[-\frac{D'''(z)}{D'(z)}+3(\frac{D''(z)}{D'(z)})^2 ] - 2 (1+z)\frac{D''(z)}{D'(z)} +1
 \end{eqnarray}
 and as a function of $H(z)$,
    \begin{eqnarray}\label{eq:q-j-h}
&q(z)& = (1+z)\frac{H'(z)}{H(z)} - 1,\\
&j(z)& = (1+z)^2[\frac{H''(z)}{H(z)}+(\frac{H'(z)}{H(z)})^2] - 2(1+z)\frac{H'(z)}{H(z)}  + 1.
\end{eqnarray}


In the case of $\Lambda$CDM model, namely 
$H(z)=H_{0}E(z)=H_{0}[\Omega_{m0}(1+z)^{3}+\Omega_{\Lambda0}]^{1/2}$ 
the cosmokinetic parameters become 
$q_{\Lambda}(z)=\frac{3}{2}\Omega_{m}(z)-1$ and $j_{\Lambda}(z)=1$, where
$\Omega(z)=\Omega_{0}(1+z)^{3}/E(z)^{2}$ and 
$\Omega_{m0}+\Omega_{\Lambda0}=1$
\footnote{For the $\Lambda$CDM model we utilize 
$\Omega_{m0}=0.3$.}. 

  
Lastly, we remind the reader the basic steps of our method (see section II). 
First the normalized distances $D(z)$ data are given as 
input to the GAPP code \cite{Seikel:2012uu}. Second we reconstruct the 
functional form of 
$D(z)$ and finally we compute the rest of the cosmological 
quantities. 
During the process we consider that the aforementioned data-sets can
be treated  as  statistically independent  measurements.
This assumption is a rather strong statement given that for example
the SNIa, QSO and GRB data are sensitive to 
luminosity distances and there might
be  spatial  overlap  between  the  various probes,  hence  this  could
lead to correlations that might affect the statistical analysis.  While
this is an important point, unfortunately at the moment  
there  is  no  standard  way  to  account  for  it  given
the lack of the full correlation matrix among 
the different samples. 
Therefore, following standard lines 
we have assumed that the different  data-sets  are  uncorrelated.
Within this framework, the corresponding parameter space is given by
$(H_{0},q_{0},j_{0})$.

\section{Results and discussion}\label{sec:res}
\begin{figure}
	\centering
	\includegraphics[width=.48\textwidth]{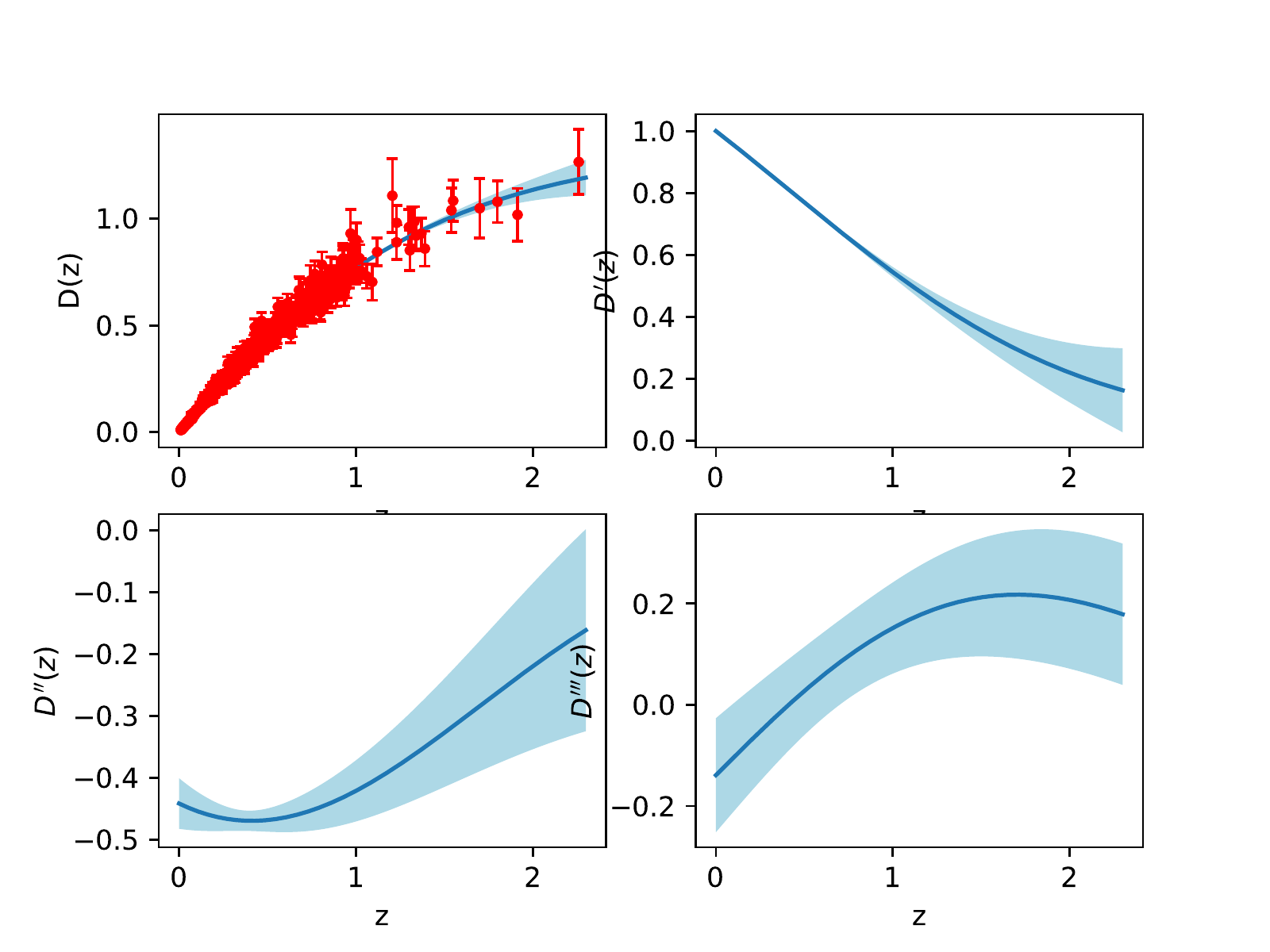}\includegraphics[width=.48\textwidth]{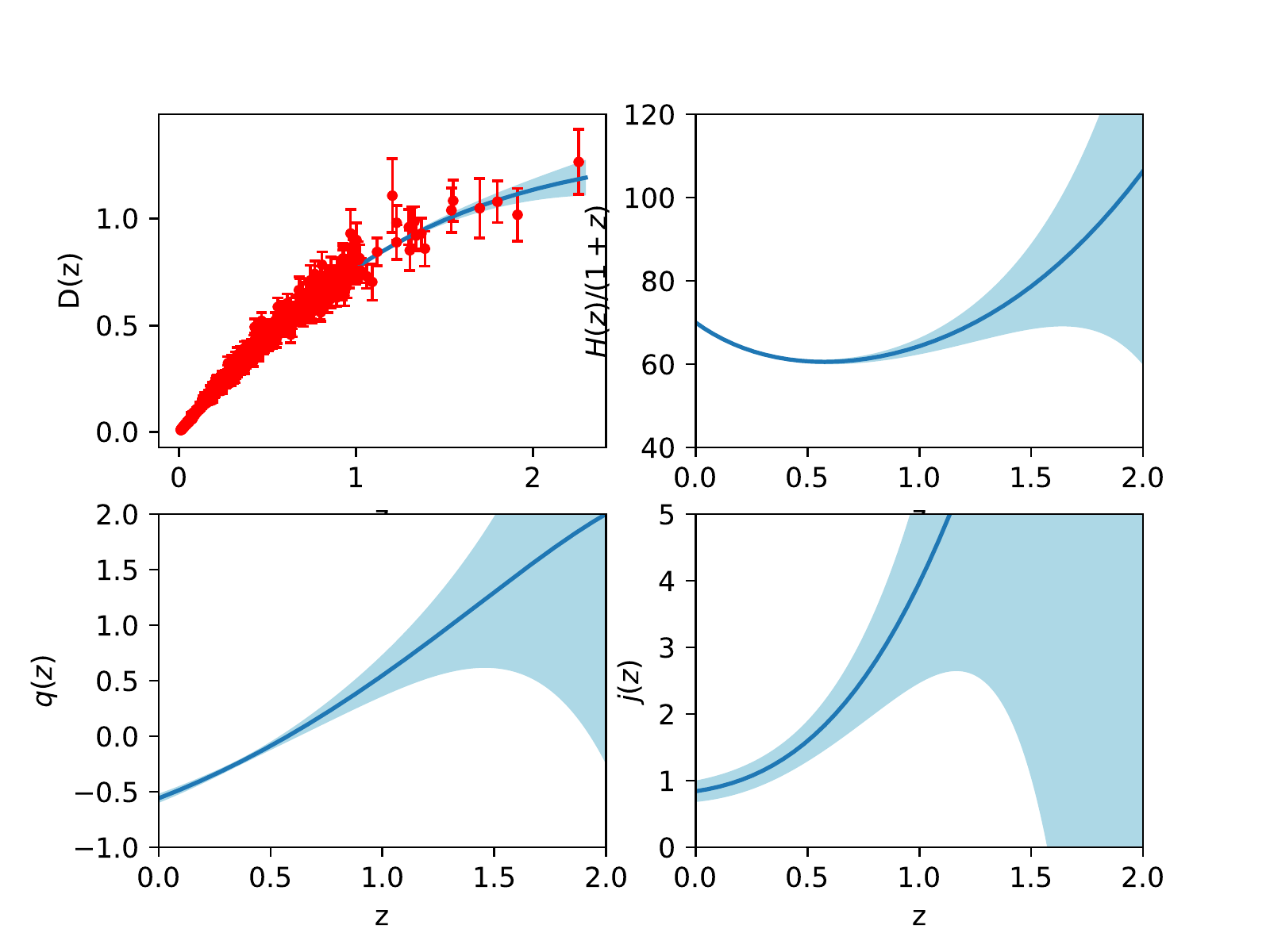}
	\caption{Left panel: Reconstruction of $D(z)$, data points and its first, second and third derivatives. Right panel: Reconstruction of $D(z)$, the Hubble function, deceleration and jerk parameters as a function of redshift for SNIa data only with the Gaussian kernel.}
	\label{fig:sn-gu}
\end{figure}

\begin{figure}
	\centering
	\includegraphics[width=.48\textwidth]{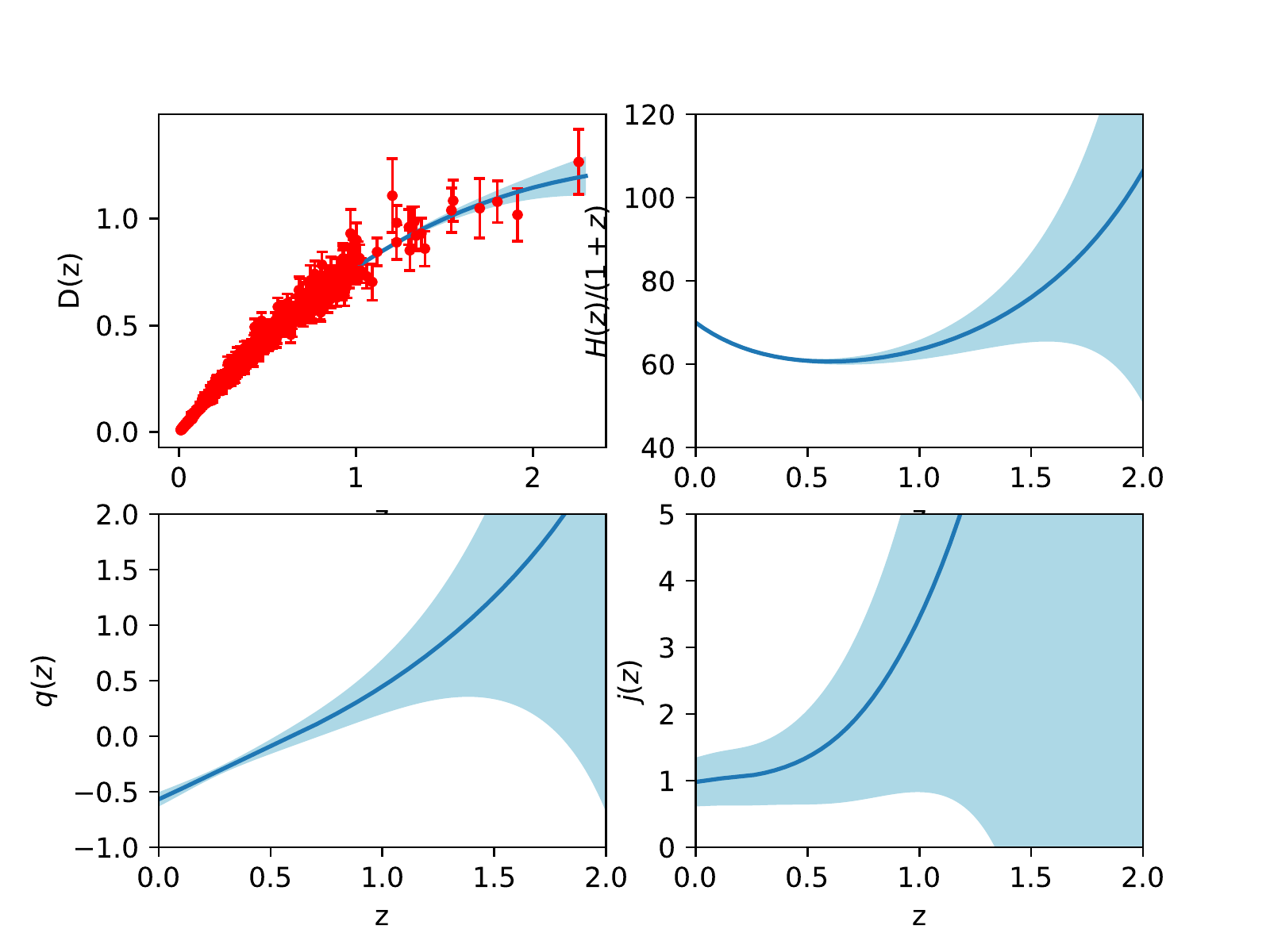}\includegraphics[width=.48\textwidth]{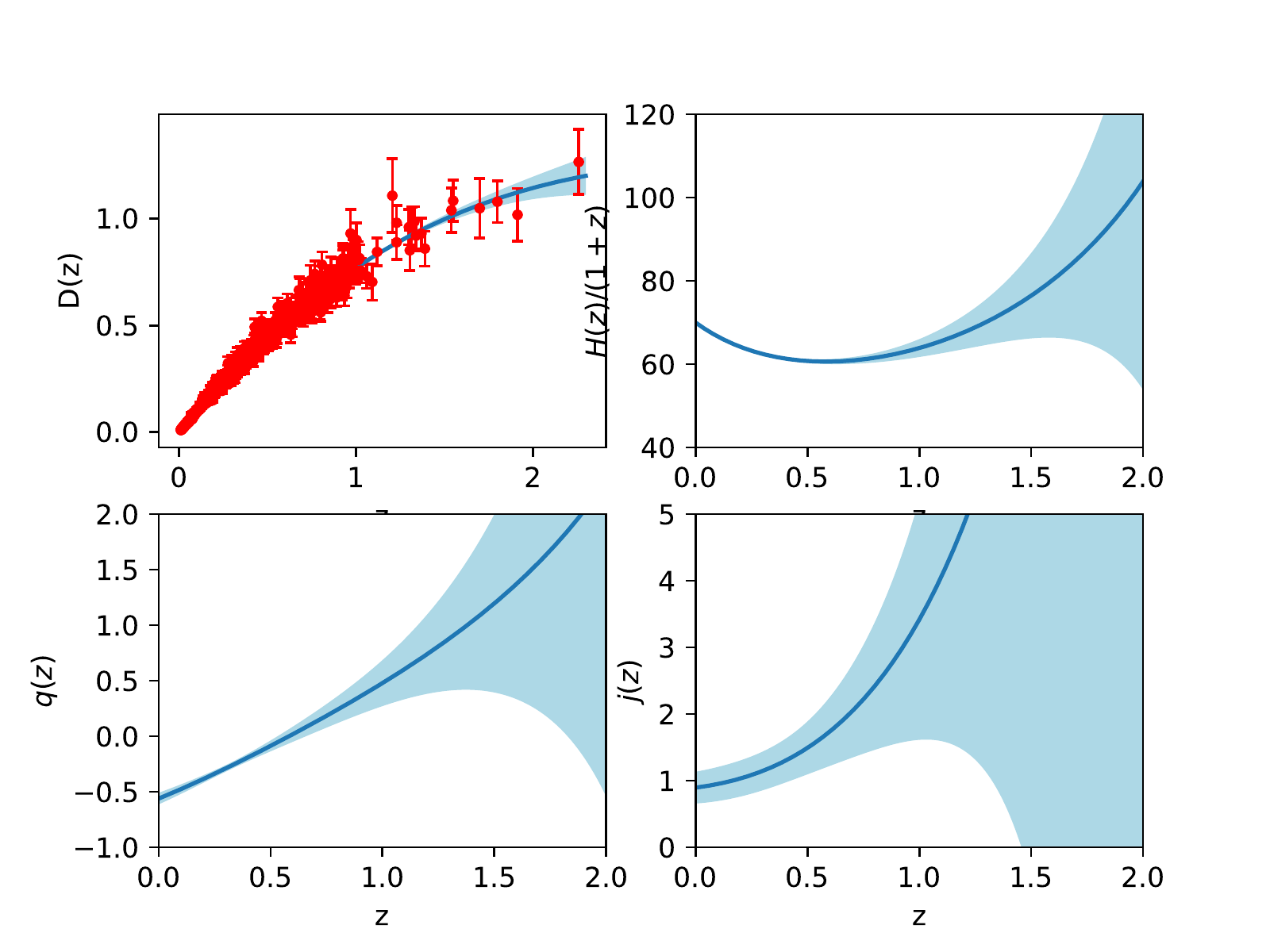}
	\caption{Left panel: The cosmokinetic parameters as a function of redshift using Matern ($\nu=7/2$) kernel for SNIa only data. 
Right panel: The cosmokinetic parameters as a function of redshift  using 
Matern ($\nu=9/2$) kernel for SN only data.}
\label{fig:sn-m}
\end{figure}

In this section, we discuss the main results of our analysis.
Specifically, in 
Table (\ref{tab:res}) we provide an overall presentation of the
cosmographic parameters at the present epoch.
In the left panel of Fig.(\ref{fig:sn-gu}) we present 
the evolution of the reconstructed $D(z)$ 
and its derivatives when using the Gaussian kernel and SNIa data. 
As expected $D^{\prime}(z)$ decreases as a function of $z$, hence due to 
Eq.(\ref{eq:conv-H}) the Hubble parameter is an increasing function.
In the right panel of Fig.(\ref{fig:sn-gu}) we plot the cosmokinetic parameters 
$H(z)/(1+z)$, $q(z)$ and $j(z)$ as a function of redshift.
Moreover in the case of Marten kernels $\nu=7/2$ and  
$\nu=9/2$ the aforementioned parameters  
are shown in Figs.(\ref{fig:sn-m}) and (\ref{fig:sn-m}) respectively.
We observe that the evolution of the kinetic parameters 
are almost the same with those of Gaussian kernel.

\begin{figure}
	\centering
	\includegraphics[width=.48\textwidth]{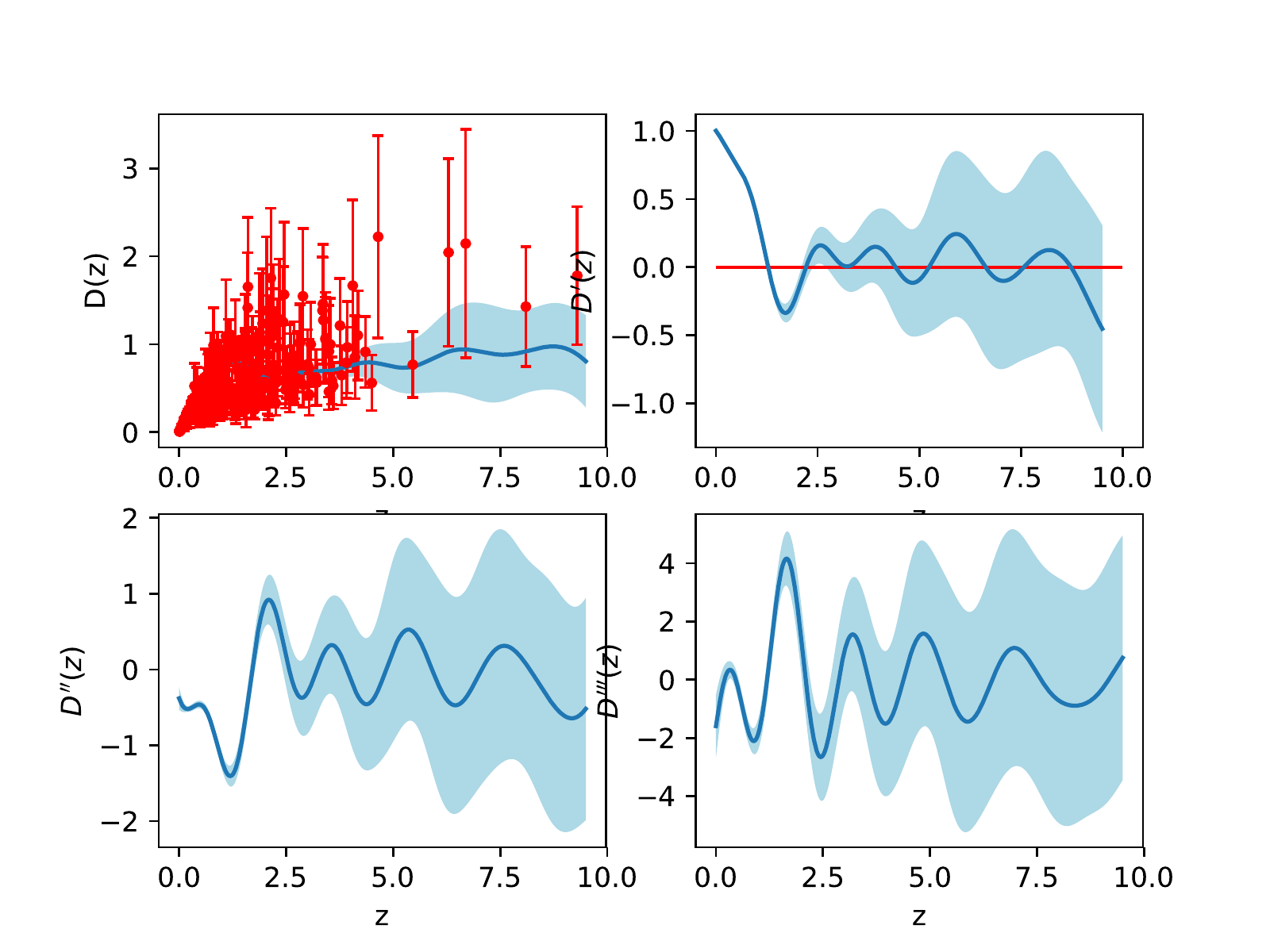}\includegraphics[width=.48\textwidth]{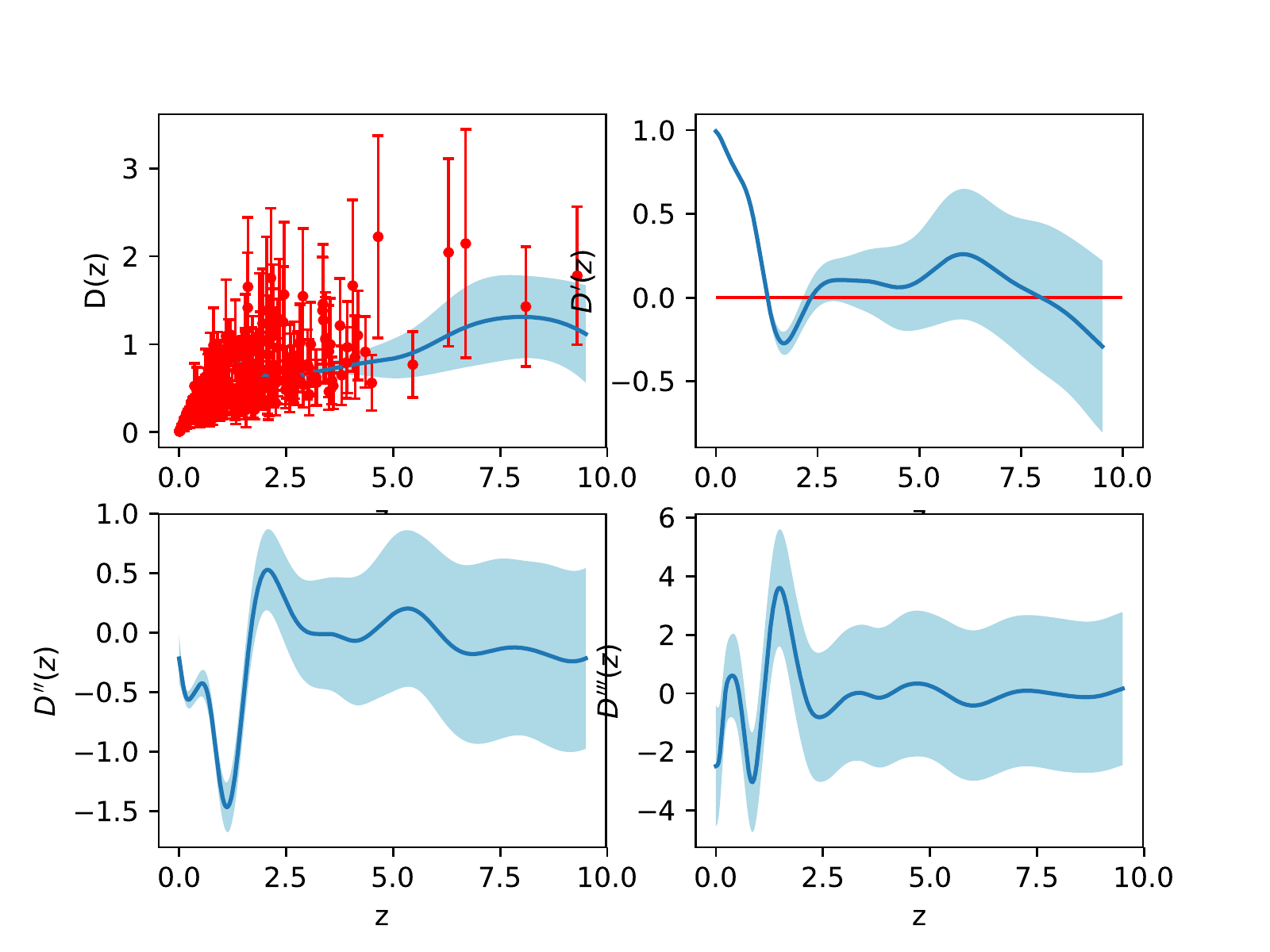}
	\caption{Left panel: The reconstruction of $D(z)$ and its derivatives as a function of redshift using Gaussian kernel by considering SNIa and GRBs. 
Right panel: The same plot using Matern ($\nu=7/2$) kernel and the same data set.}
\label{fig:sn-gr}
\end{figure}

However, when combined SNIa with other probes, such as GRBs, 
the situation becomes different. Indeed, for SNIa/GRBs we plot  
in Fig.(\ref{fig:sn-gr}) 
$D(z)$ and its derivatives versus redshift
using the Gaussian (left panel) and Matern $\nu=7/2$ (right panel) kernels.  
For both cases we observe that in the evolution of the 
corresponding derivatives appears oscillations. It is easy to check that  
the first derivative of $D(z)$ crosses the zero line several times, hence 
the cosmokinetic parameters diverge at these points. 
Notice that utilizing the Matern $\nu=9/2$ kernel the results remain unaltered.
We argue that although GRBs may help to reconstruct the cosmic expansion 
up to $z\sim 10$, however there are 
practical difficulties in achieving this goal in the case of Gaussian process.

\begin{figure}
	\centering
	\includegraphics[width=.48\textwidth]{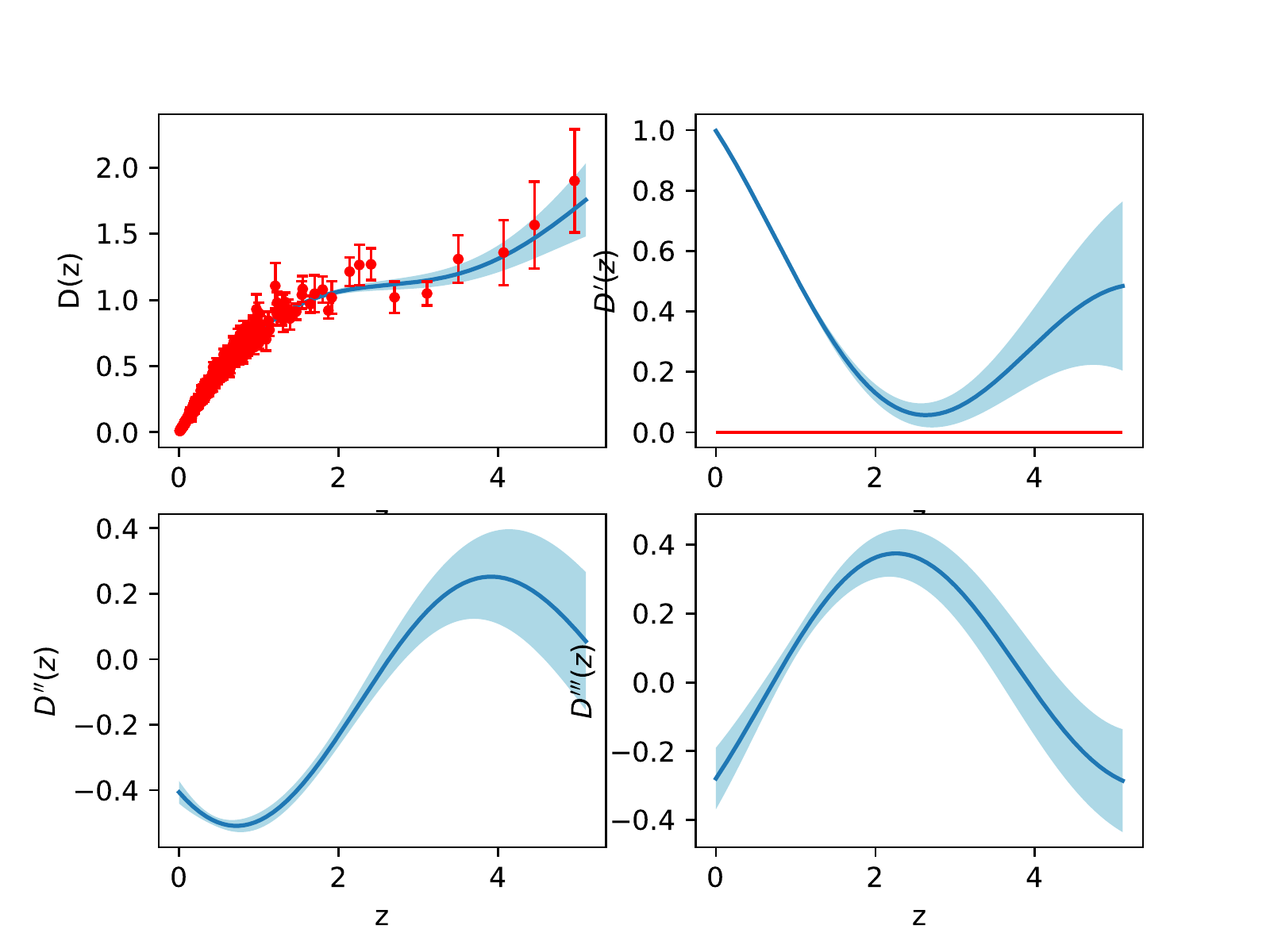}\includegraphics[width=.48\textwidth]{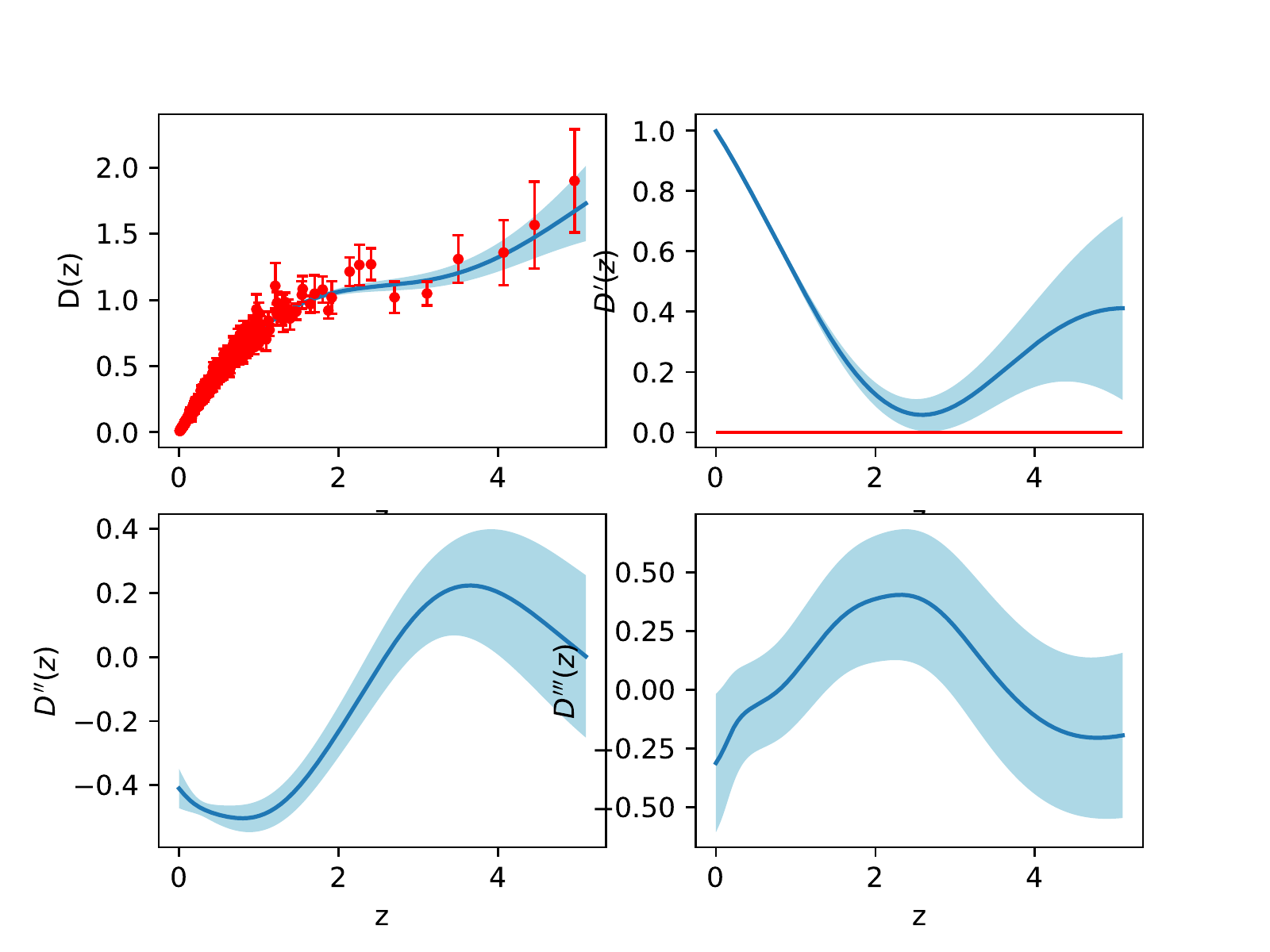}
	\caption{Left panel: The reconstruction of $D(z)$ and its derivatives as a function of redshift using Gaussian kernel by considering SNIa and QSOs. Right panel: The same plot using Matern ($\nu=7/2$) kernel and the same data set.}
	\label{fig:sn-qa}
\end{figure}

Moreover, the results of SNIa/QSO combination 
are presented in Fig.(\ref{fig:sn-qa}) 
and Tab.(\ref{tab:res}). 
In this case, we observe that $D(z)$ slowly decreases prior to $z\sim 3$,
hence a small oscillation appears at that redshift.
Furthermore, we find that both Gaussian and Matern kernels provide 
similar results and in contrast to SNIa/GRB case, here
the first derivative of the $D(z)$ does not cross 
the zero line at 1 $\sigma$ level.       

\begin{figure}
	\centering
	\includegraphics[width=.48\textwidth]{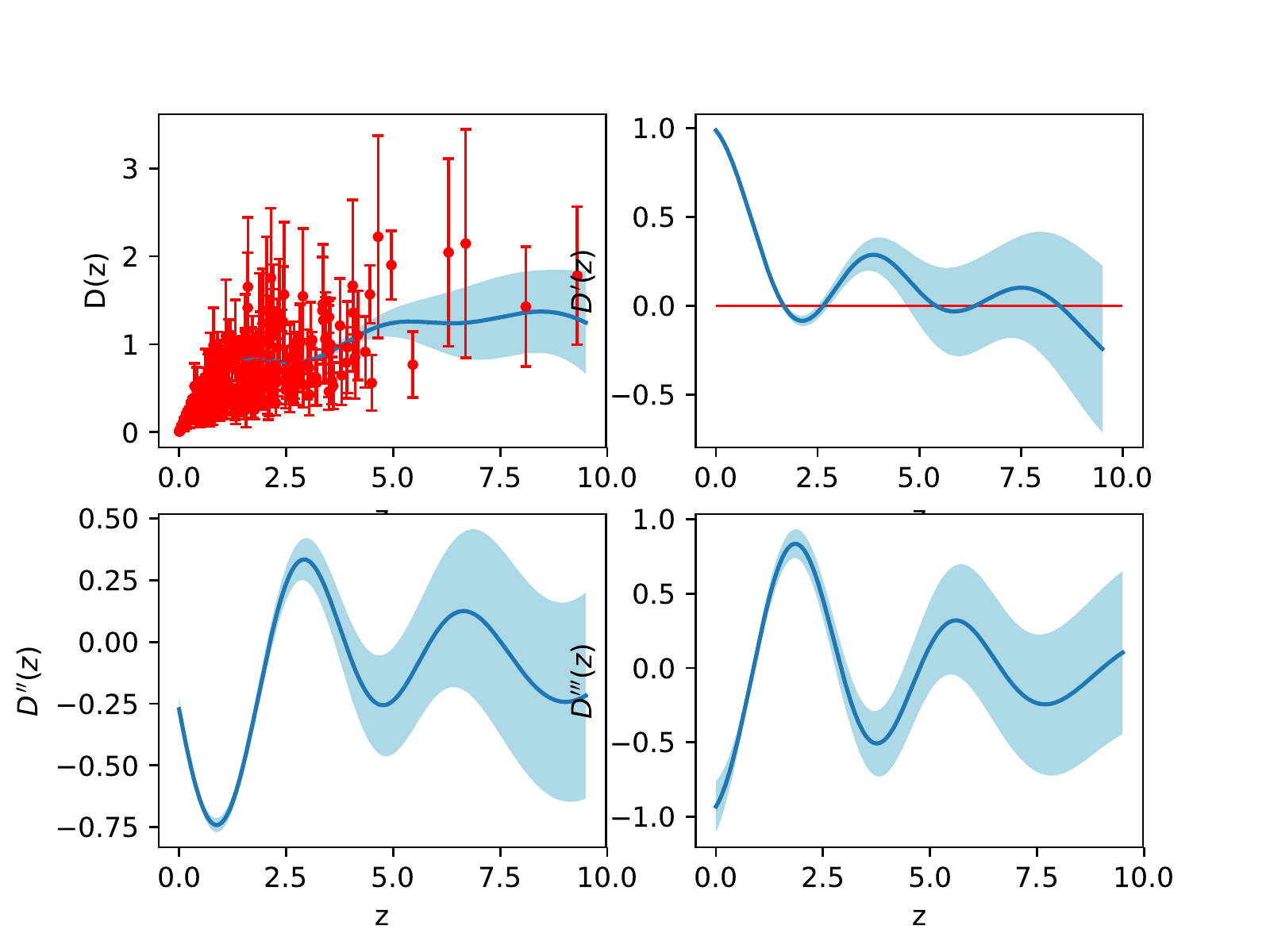}\includegraphics[width=.48\textwidth]{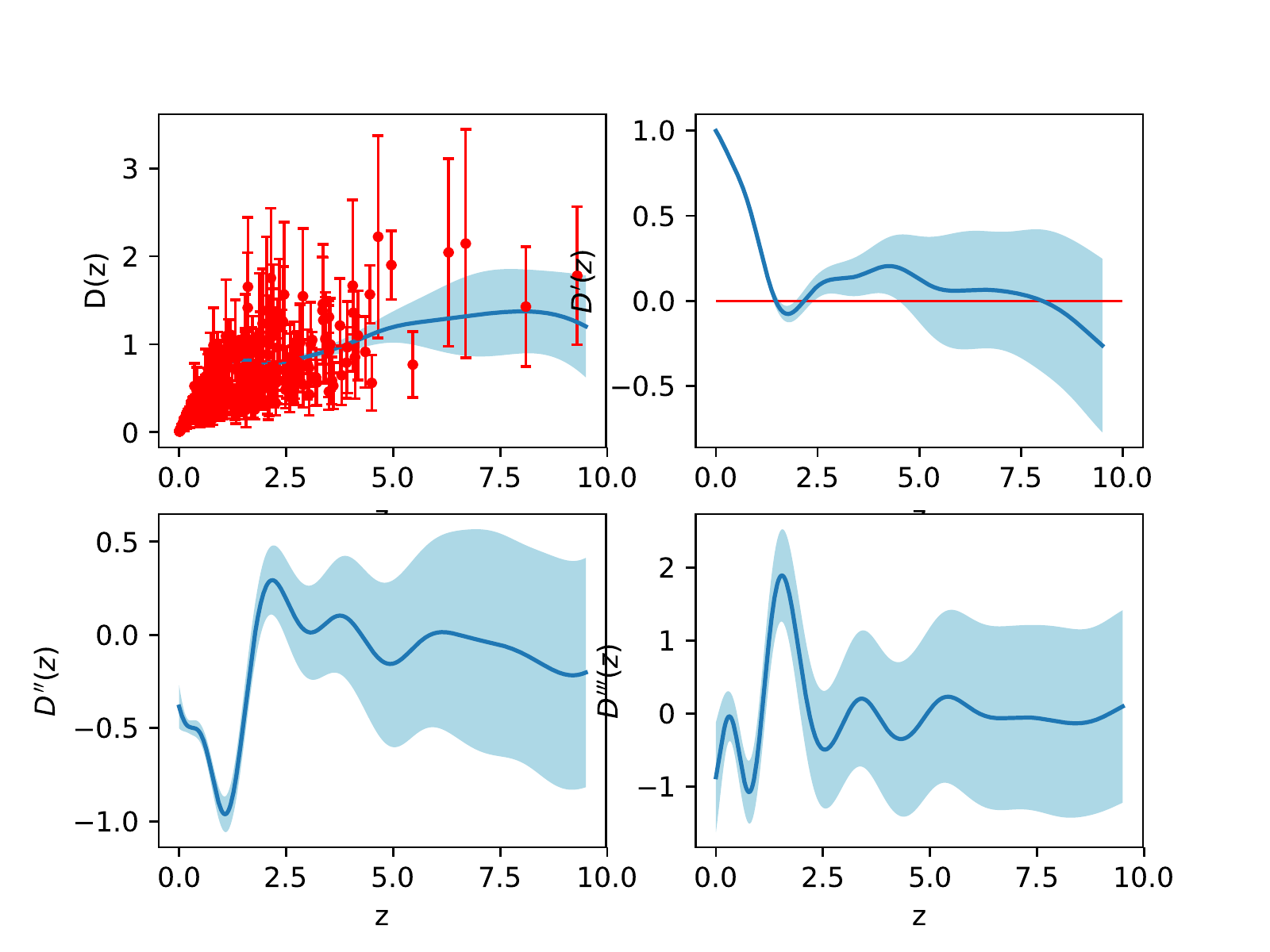}
	\caption{Left panel: The reconstruction of $D(z)$ and its derivatives as a function of redshift using Gaussian kernel for all data set. Right panel: The same plot using Matern ($\nu=7/2$) kernel and the same data set.}\label{fig:sn-gr-qu}
\end{figure}

Lastly, we combine SNIa, GRBs and QSOs in order to 
compute the reconstructed comoving distance for all kernels.  
As an example in Fig.(\ref{fig:sn-gr-qu}) we plot the evolution 
of $D(z)$ and the corresponding derivatives 
in the case of Matern $\nu=7/2$ kernel.
Again we verify that there are epochs which are located 
at large redshifts and for which $D^{\prime}(z)$ 
crosses the zero line (similar behavior is found for the other kernels).


\begin{table}
	\begin{center}
		\begin{tabular}{ |c|c|c|c| } 
			\hline
			 \diagbox{Data set}{Kernel}  & Gaussian  &  Matern $\nu=7/2$&  Matern $\nu=9/2$ \\
			\hline
			SNIa & \makecell{$H_0$= $70.00\pm 0.38$ $\vspace{0.2 cm}$\\ $q_0$=$-0.558\pm 0.04$ $\vspace{0.2 cm}$ \\ $j_0$=$0.84\pm 0.16$} &  \makecell{$H_0$=$69.98\pm 0.46$ $\vspace{0.2 cm}$ \\ $q_0$=$-0.567\pm 0.06$ $\vspace{0.2 cm}$\\ $j_0$=$0.98\pm 0.36$}& \makecell{$H_0$=$69.98\pm 0.42$ $\vspace{0.2 cm}$ \\ $q_0$=$-0.561\pm 0.05$ $\vspace{0.2 cm}$\\ $j_0$=$0.89\pm 0.23$} \\
			\hline
			SNIa+GRBs & \makecell{$H_0$=$69.92\pm 0.72$  $\vspace{0.2 cm}$\\ $q_0$=$-0.62\pm 0.15$ $\vspace{0.2 cm}$\\ $j_0$=$2.26\pm 1.1$} &  \makecell{$H_0$=$70.49\pm 0.85$ $\vspace{0.2 cm}$ \\ $q_0$=$-0.79\pm 0.20$ $\vspace{0.2 cm}$\\ $j_0$=$3.21\pm 2.1$}& \makecell{$H_0$=$70.28 \pm 0.77$ $\vspace{0.2 cm}$ \\ $q_0$=$-0.73\pm 0.17$ $\vspace{0.2 cm}$\\ $j_0$=$2.80\pm1.4$} \\
			\hline
			SNIa+QSOs & \makecell{$H_0$=$70.17\pm 0.35$  $\vspace{0.2 cm}$\\ $q_0$=$-0.59\pm 0.03$ $\vspace{0.2 cm}$\\ $j_0$=$0.96\pm 0.13$} &  \makecell{$H_0$=$70.13\pm 0.45$ $\vspace{0.2 cm}$ \\ $q_0$=$-0.58\pm 0.06$ $\vspace{0.2 cm}$\\ $j_0$=$0.99\pm 0.33$}
			& \makecell{$H_0$=$70.15 \pm 0.40$ $\vspace{0.2 cm}$ \\ $q_0$=$-0.59\pm 0.05$ $\vspace{0.2 cm}$\\ $j_0$=$0.95\pm0.21$} \\
			\hline
			SNIa+GRBs+QSOs & \makecell{$H_0$=$70.86\pm 0.42$ $\vspace{0.2 cm}$ \\ $q_0$=$-0.72\pm0.05$ $\vspace{0.2 cm}$\\ $j_0$=$1.62\pm0.2$} &  \makecell{$H_0$=$70.22\pm0.69$ $\vspace{0.2 cm}$ \\ $q_0$=$-0.66\pm0.14$ $\vspace{0.2 cm}$\\ $j_0$=$1.98\pm1.2$}& \makecell{$H_0$=$70.12\pm 0.63$ $\vspace{0.2 cm}$ \\ $q_0$=$-0.62\pm0.11$ $\vspace{0.2 cm}$\\ $j_0$=$1.55\pm 0.81$} \\
			\hline
		\end{tabular}
	\end{center}
	\caption{Cosmokinetic parameters at present time for different data sets and kernels. }\label{tab:res} 
\end{table}  

Now we focus on Tab. (\ref{tab:res}) which shows the 
cosmokinetic parameters at the present time for various data and kernels
explored in this study.
Considering only the traditional standard candles (SNIa), we find that 
the Hubble constant is close to ~70${\rm Km/sec/Mpc}$ regardless the 
form of kernel, while 
$q_0$ and $j_0$ are consistent (within $1\sigma$) with those of $\Lambda$CDM. 
Combining SNIa and GRB data, we find that the value of $H_0$ 
does not change significantly and it remains close to 
~70${\rm Km/sec/Mpc}$. In the case of Gaussian kernel, the current value of
the deceleration parameter is in agreement with that of 
$\Lambda$CDM at 1 $\sigma$ level.  
For the Matern's kernels the extracted value 
of $q_{0}$ is marginally consistent with $\Lambda$CDM with $q_{0}<q_{\Lambda,0}$.
Concerning $j_{0}$, our results are similar to those 
of \cite{Lusso:2019akb}, however the corresponding uncertainties 
are larger (by a factor of 2.5-4) 
than those of \cite{Lusso:2019akb}, implying that the extracted jerk parameters
are consistent with the the predictions of $\Lambda$CDM at
 $~ 2\sigma$ level. 
Combining SNIa and QSO datasets, we find that for all kernels the cosmokinetic
parameters $(q_{0},j_{0})$ are in a good agreement (with $1\sigma$) with those 
of $\Lambda$CDM model. 

 \begin{figure}
	\centering
	\includegraphics[width=.48\textwidth]{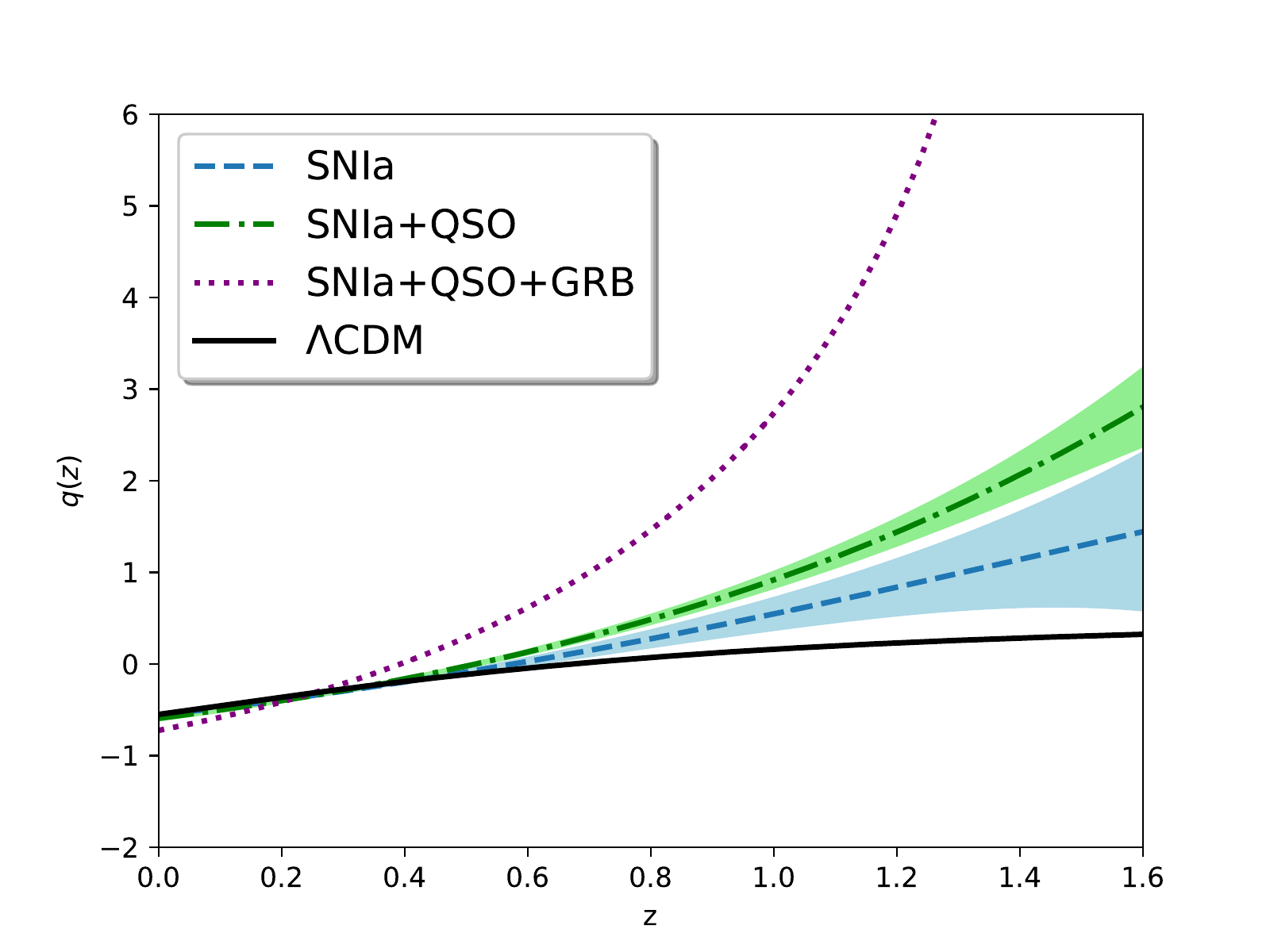}\includegraphics[width=.48\textwidth]{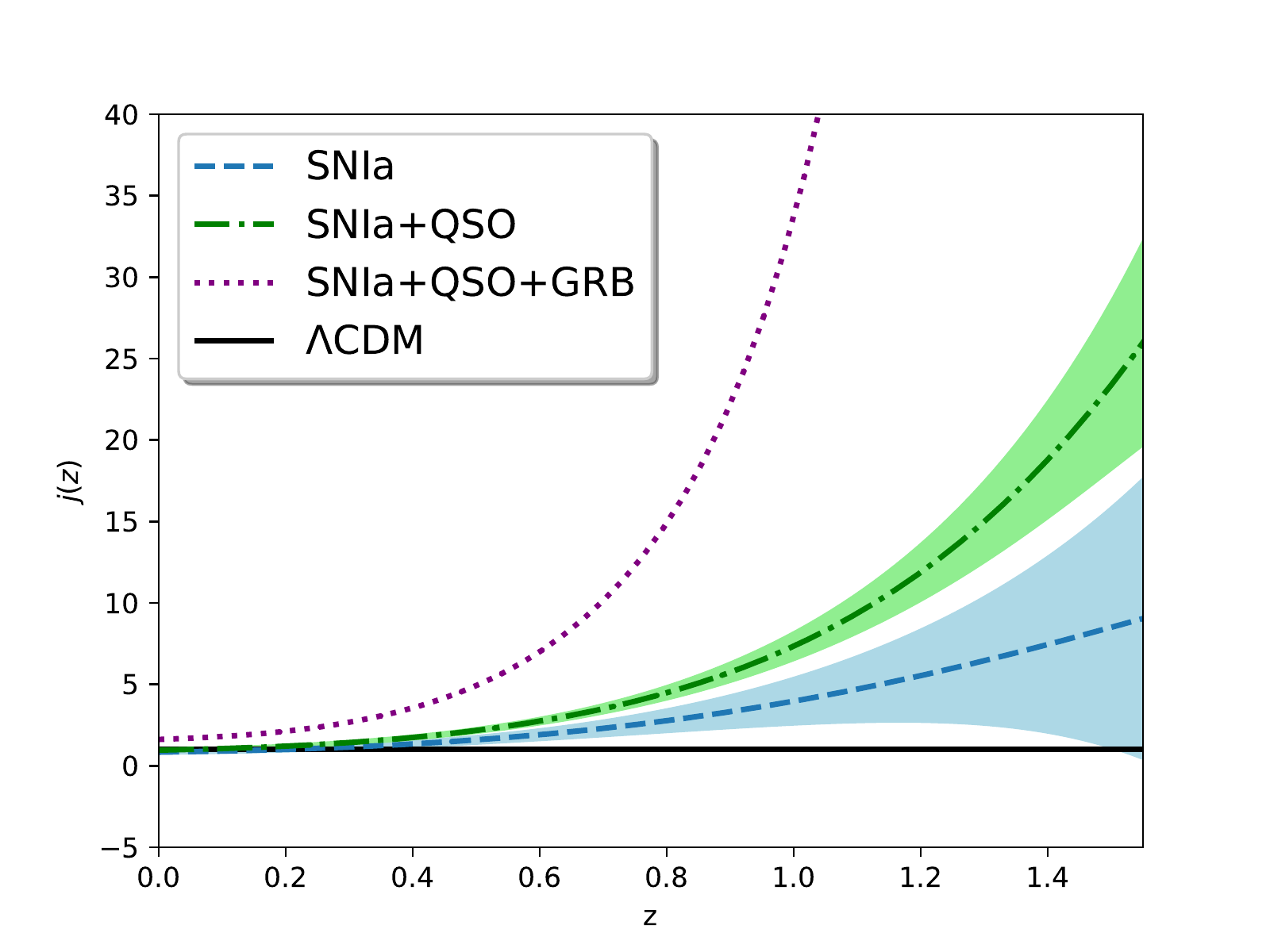}
	\caption{Left panel: Reconstruction of $q(z)$ using different data sets and considering the Gaussian kernel. Right panel: Reconstruction 
of $j(z)$. In the case of $\Lambda$CDM model we use $\Omega_{m0}=0.3$ 
(see solid black lines).}
	\label{fig:q-j}
\end{figure}

Finally, in the case of the Gaussian kernel 
the combination SNIa/QSOs/GRBs indicates 
that the extracted values of $q_0$ and $j_0$ are $\sim 3\sigma$ away 
from those of $\Lambda$CDM. However, the opposite situation 
holds in the case of Matern's kernels, namely both $q_0$ and $j_0$ are 
consistent (due to large uncertainties) with the predictions of $\Lambda$CDM. 
In a nutshell, for the usual standard candles (SNIa data) 
and for the combination SNIa/QSOs 
we find that the cosmokinetic parameters $(q_{0},j_{0})$ extracted from the  
Gaussian process are consistent with $\Lambda$CDM. 
However, including 
GRBs in the analysis we find 
a tension of the $\Lambda$CDM model which lies between 
2 and 3 $\sigma$ levels respectively.
Moreover, the combined SNIa/QSO/GRB analysis shows that the choice of 
the kernel function might affect the amount of tension. 
Indeed in the case of Matern's kernels we produce cosmokinetic parameters 
which are consistent with those of $\Lambda$CDM, while using 
the Gaussian kernel it seems that the $\Lambda$CDM model is in tension 
with the measurements $(q_{0},j_{0})$.

\subsection{Cosmokinetic parameters at high redshits}
Apart from $(q_{0},j_{0})$ it is useful to 
study the cosmokinetics parameters at high redshifts. 
For the Gaussian kernel we plot in Fig.(\ref{fig:q-j}) the evolution 
of $q(z)$ and $j(z)$ in the case of SNIa (blue dashed line), SNIa/QSO 
(green dot-dashed) and SNIa/QSO/GRBs (magenta dotted curve).
For comparison we also plot $q_{\Lambda}(z)$ and $j_{\Lambda}(z)$
(see solid lines). Since $D^{'}(z)$ may
cross the zero line prior to $z\sim 2$ we prefer to focus 
on $1<z<2$. Obviously, a strong deviation from the $\Lambda$CDM predictions
is observed in the case 
of SNIa/QSO and SNIa/QSO/GRBs. We also checked that this result persists 
regardless the form of the kernel.
Although the situation regarding the cosmokinetic 
tension is not so clear in the present epoch, at high redshifts 
there is a clear indication that such a tension really exists.
Especially, the jerk parameter clearly points to this direction, hence 
the possibility of having new Physics is not excluded
by the present analysis. Notice that, our results are in agreement with those of
\cite{Lusso:2019akb,Risaliti:2018reu} who found that the deviation 
from the flat $\Lambda$CDM becomes strong at high redshifts ($z>1)$.
Combining our model-independent parametrization of the Hubble Diagram 
with those of \cite{Lusso:2019akb,Risaliti:2018reu} we conclude 
that the deviation from the concordance $\Lambda$CDM 
model is due to new Physics.



\section{Conclusion}\label{sec:con}
It is well known that the concordance 
$\Lambda$CDM model fits accurately the current
cosmological data \cite{Aghanim:2018eyx}, nonetheless it has been proposed 
that the model is not without its problems. Indeed there are indications 
that the $\Lambda$CDM model is in tension with some 
important measurements~\cite{Verde_2019,Sola:2017znb}, 
namely the Hubble constant $H_0$ and 
the present value of the mass variance at 8$h^{-1}$Mpc, namely $\sigma_8$. 
In this context, Lusso {\it et al.} \cite{Lusso:2019akb} using 
a combined Hubble diagram of SNIa, Quasars, and Gamma-Ray Bursts (GRBs) 
found a $\sim 4\sigma$ tension between the 
best fit cosmokinetic parameters with respect to those of $\Lambda$CDM
(see also \cite{Risaliti:2018reu}).  
Whether the above tensions are the result of yet unknown 
systematic errors or indicate some underlying
new Physics is still an open issue. Therefore, 
on this subject an intense debate is taking place in the literature
and the aim of the present work is to contribute to this debate.

In particular, we combined the traditional standard candles
(SNIa data) with other extragalactic sources (Quasars and GRBs) 
to reconstruct, in a model independent way, the Hubble diagram 
to as high redshifts as possible and to compute the corresponding
cosmokinetic parameters at the present epoch, namely deceleration $q_{0}$ and 
jerk $j_{0}$ parameters.
Using only the SNIa data 
we found that the cosmokinetic parameters $(q_{0},j_{0})$ extracted from the  
Gaussian process are consistent with those of $\Lambda$CDM. 
Also in the case of SNIa/QSO combination, we found 
that for all kernels the cosmokinetic
parameters are in a very good agreement (with $1\sigma$) with those 
of $\Lambda$CDM model. 


 On the other hand combining SNIa with Quasars and GRBs 
we revealed some tension, which lies between
$2\sigma$ and $3\sigma$ levels, depending on the kernel choice.
Finally, focusing our analysis on high redshifts ($z>1$)  
we found that the corresponding cosmokinetic parameters 
significantly deviate from those of $\Lambda$CDM. Overall the combination 
of the present work with those of \cite{Lusso:2019akb,Risaliti:2018reu}
provide a complete investigation of the so called 
$\Lambda$CDM ''tension''. The three works, which are model independent,
clearly suggest that the discrepancy between the Hubble diagram data 
(especially for $z>1$) and the predictions of the concordance $\Lambda$CDM 
model is the result of some underlying new Physics.

  \bibliographystyle{apsrev4-1}
\bibliography{ref}

\end{document}